\documentclass[aps,prd,10pt,twocolumn,nofootinbib,floatfix,
  noshowpacs,preprintnumbers,superscriptaddress]{revtex4-1}
\usepackage{bm}
\usepackage{epsfig}
\usepackage{amsmath}
\usepackage{tablefootnote}
\usepackage{color}
\usepackage[dvipsnames]{xcolor}
\usepackage[colorlinks=true,breaklinks=true]{hyperref}
\hypersetup{allcolors=[rgb]{0.0 0.0 0.6},linkcolor=[rgb]{0.75 0.05 0.05}}
\usepackage[normalem]{ulem}

\newcommand{\exclude}[1]{}

\begin{document}

\title{Extending the Reach of Leptophilic Boson Searches at DUNE and MiniBooNE with Bremsstrahlung and Resonant Production}

\author{Francesco Capozzi}  
\affiliation{Center for Neutrino Physics, Department of Physics, Virginia Tech, Blacksburg, VA 24061, USA}

\author{Bhaskar Dutta}
\affiliation{Mitchell Institute for Fundamental Physics and Astronomy, Department of Physics and Astronomy, Texas A\&M University, College Station, TX 77845, USA}

\author{Gajendra Gurung}
\affiliation{Department of Physics, University of Texas, Arlington, TX 76019, USA}  

\author{Wooyoung Jang}
\affiliation{Department of Physics, University of Texas, Arlington, TX 76019, USA}  

\author{Ian M. Shoemaker}
\affiliation{Center for Neutrino Physics, Department of Physics, Virginia Tech, Blacksburg, VA 24061, USA}

\author{Adrian Thompson}
\affiliation{Mitchell Institute for Fundamental Physics and Astronomy, Department of Physics and Astronomy, Texas A\&M University, College Station, TX 77845, USA}

\author{Jaehoon Yu}
\affiliation{Department of Physics, University of Texas, Arlington, TX 76019, USA}  

\begin{abstract}
New gauge bosons coupling to leptons are simple and well-motivated extensions of the Standard Model. We study the sensitivity to gauged $L_{\mu} -L_{e}$, $L_e-L_\tau$ and $L_{\mu} -L_{\tau}$ both with the existing beam dump mode data of MiniBooNE and with the DUNE near detector. We find that including bremsstrahlung and resonant production of $Z'$ which decays to $e^{\pm}$ and $\mu^{\pm}$ final states leads to a significant improvement in existing bounds, especially for $L_\mu-L_e$ and $L_e-L_\tau$ for DUNE while competitive constraints can be achieved with the existing data from the MiniBooNE's beam dump run.

\end{abstract}
\preprint{MI-HET-752}
\maketitle

\section{Introduction}

Despite the presence of several hints in favor of the existence of physics beyond the Standard Model (BSM), such as neutrino masses, dark matter and dark energy, the scale at which this new physics resides is unknown. The searches at neutrino experiments with high intensity beams offer a possibility of finding new physics at relatively low energy scales with small couplings to the SM.

One generic possibility for BSM physics is additional gauge symmetries. Typically, new gauge symmetries come with their related anomaly cancellation requirements, implying the need for a number of new fermions. However, one class of symmetries which can be gauged without implying any new fermions are those associated with the lepton number~\cite{He:1990pn,He:1991qd}. In this case, pairs of the individual lepton flavor numbers are gauged, with each pair having an equal and opposite charge assignment, i.e. $L_{\alpha}-L_{\beta}$, where $\alpha, \beta = e,\mu,\tau$. The gauging of such symmetries only directly predicts the existence of a new $Z'$ gauge boson associated with the symmetry. The SM Lagrangian is thus extended by
\begin{equation}
{\cal L} = {\cal L}_{\rm SM} - \frac{1}{4} {Z^\prime}^{\delta\eta}Z^\prime_{\delta\eta} + \frac{m^2_{Z^\prime}}{2}  {Z^\prime_\delta}  {Z^\prime}^\delta    + Z^{\prime}_\delta J^{\delta}_{\alpha - \beta},
\label{eq:lagrangian}
\end{equation}
where $m_{Z^\prime}$ is the mass of the new gauge boson and $J^{\delta}_{\alpha - \beta}$ is a current defined as
\begin{equation}
J^{\delta}_{\alpha - \beta} = g_{\alpha\beta} \left(    \bar{l}_{\alpha} \gamma^\delta l_{\alpha} + \bar \nu_{\alpha} \gamma^\delta P_L \nu_{\alpha}  -
  \bar{l}_{\beta} \gamma^\delta l_{\beta}-  \bar \nu_{\beta} \gamma^\delta P_L \nu_{\beta}    \right),
  \label{eq:current}
\end{equation}
with $g_{\alpha\beta}$ being the coupling between $Z^\prime$ and both charged leptons $l_\alpha$ and neutrinos $\nu_\alpha$. The symmetries of the model allow for kinetic mixing between the $Z^\prime$ and the SM gauge bosons. This corresponds to the following extra term of the Lagrangian
\begin{equation}
{\cal L}_{\epsilon} = -\frac{\epsilon}{2}F_{\gamma\delta}Z^{\prime\gamma\delta}\,,
\label{lagrangian_kinetic_mixing}
\end{equation}
where $F_{\gamma\delta}$ is field strength of hypercharge. Unless explicitly specified, in this work we assume a vanishing tree level contribution to $\epsilon$, which is then only generated at the loop level, and it is given by
 \begin{equation}
\epsilon(q^2)=\frac{e\,g_{\alpha\beta}}{2\pi^2}\int_0^1dx\,x(1-x)\log\frac{m^2_\alpha-x(1-x)q^2}{m^2_\beta-x(1-x)q^2}\,,
\label{kinetic_mixing}
\end{equation}
where $e$ is the electric charge, $m_{e,\mu,\tau}$ is the mass
of a charged lepton, and $q^2$ is the momentum transfer. 

The presence of leptophilic gauge bosons has been extensively looked for in lepton colliders or fixed target beam dump experiments, as well as in neutrino experiments (see \cite{Bauer:2018onh} for a detailed review of the current experimental bounds). In general, for $m_{Z^\prime}\in[1,10^3]$ MeV and $g_{\alpha\beta}<10^{-5}$ the most stringent constraints come from electron beam dump experiments, such as E137 \cite{Bjorken:1988as}, with a worse sensitivity for $L_\mu-L_\tau$ because of the coupling to electrons being possible only at the loop level. It has recently been shown in \cite{Berryman:2019dme} that DUNE \cite{DUNE:2020lwj,DUNE:2020ypp,DUNE:2020mra,DUNE:2020txw} might play a role in constraining $L_{\alpha}-L_{\beta}$ models. 

In particular, it has been pointed out that the charged mesons created in proton beam interactions can produce a relatively large number of $Z^\prime$, whose subsequent decay into lepton pairs induces a sizeable signal in the detector. This will improve current constraints, but only for $L_{\mu}-L_{\tau}$ and only in a narrow window of masses around $5-10$ MeV. At the same time, in \cite{Nardi:2018cxi,Celentano:2020vtu} the relevance as production channels of $e^\pm$ bremsstrahlung, as well as of on-shell resonance has been noted in the context of dark photon models. Despite the fact that such channels can dominate over charged meson decays, especially in the context of $L_\mu-L_e$ and $L_e-L_\tau$, they have not been included in \cite{Berryman:2019dme}. On the other hand in \cite{Nardi:2018cxi,Celentano:2020vtu} the main focus is on a generic kinetic mixing model in the context of light dark matter. 

The purpose of our work is to show that the $Z^\prime$ production from both bremstrahlung and on-shell resonance in DUNE can significantly improve the prospective sensitivity =to $L_\alpha-L_\beta$ gauge boson. Another result of our work is that the existing data collected in the beam dump mode of MiniBooNE \cite{Aguilar-Arevalo:2018wea} provides already competitive bounds. We note that experiments such as DUNE and MiniBooNE will have sensitivity to a variety of BSM physics, including heavy neutral leptons~\cite{Ballett:2019bgd,Coloma:2020lgy,Breitbach:2021gvv,Atkinson:2021rnp,Schwetz:2021crq}, axion-like particles~\cite{Kelly:2020dda,Brdar:2020dpr}, dark matter~\cite{Batell:2009di,deNiverville:2012ij,Coloma:2015pih,deNiverville:2018dbu,Jordan:2018qiy,deGouvea:2018cfv,DeRomeri:2019kic,Breitbach:2021gvv}, and millicharged particles~\cite{Magill:2018tbb}. Finally, we note that additional, complementary signatures of $Z'$s can arise from modifications to the neutrino scattering cross section~\cite{Ballett:2019xoj,Dev:2021xzd}. 

The structure of the paper is as follows. In Section~\ref{sec:GEANT}, we briefly outline the physics models used in the GEANT4 simulation of the proton beam interactions with both the graphite target in DUNE and the beam dump in MiniBooNE with their material and geometries implemented. 
In Section~\ref{sec:production}, we describe how we convert the products of proton interactions, e.g. photons, into a corresponding number of $Z^\prime$ entering the neutrino detectors. 
In Section~\ref{sec:detector}, we present the mathematical framework for calculating the expected number of lepton pairs produced by $Z^\prime$ decays in a detector. 
In Section~\ref{sec:sens}, we perform an estimate of the sensitivity to the $m_{Z^\prime}$ and $g_{\alpha\beta}$ of both MiniBooNE and DUNE. A direct comparison with the results and simulations obtained in  \cite{Berryman:2019dme,Celentano:2020vtu} is also presented, along with constraints from astrophysical observations and electron beam dump experiments. 
In Section~\ref{sec:cons}, we discuss the detailed assumptions underlying these constraints and compare them to the MiniBooNE and DUNE sensitivity obtained in this work.

\section{GEANT4 Simulation}
\label{sec:GEANT}
The neutrinos in MiniBooNE and DUNE are produced from the decays of the secondary mesons generated by the protons interactions in the target, as in all neutrino experiments. 
In order to simulate the production of all particles in the proton beam interactions, we used GEANT4 simulation tool kit~\cite{GEANT4:2002zbu,Allison:2006ve,Allison:2016lfl}. 
For the MiniBooNE beam dump mode, we implemented the target, namely the beam dump geometry as described in Ref.~\cite{PhysRevD.98.112004}. 
The MiniBooNE dump dimension is 4 m (width) $\times$ 4 m (height) $\times$ 4.21 m (length) of which the upstream most 2.64 m is stainless steel, followed by a 0.91 m thick concrete and finally a 0.66 m thick  stainless steel layers along the beam direction.
The implemented DUNE neutrino production target is a 1.5 m long cylindrical graphite rod with 1.7~cm diameter, following the description in the LBNF beamline design \cite{Papadimitriou:2017zai}.

For both MiniBooNE and DUNE simulations, we used \texttt{QGSP\_BIC\_AllHP} physics list for the hadronic reaction and \texttt{G4EmStandardPhysics} for the electromagnetic interactions. In addition, we have developed an inherited user-defined class of \texttt{G4UserSteppingAction}, derived from \texttt{G4SteppingAction}, to trace and record all the particles produced in the proton beam interactions as they progress throughout the target.  In particular, we recorded 4-momenta of all photons produced in the target from the primary proton interaction to electromagnetic showering process since these are used to calculate the $Z^\prime$ production, as described in the next Section.

\section{Production channels of $Z^\prime$}
\label{sec:production}
The secondary particles we are interested in are photons and positrons. In the context of $L_\alpha-L_\beta$ models, each photon can be substituted by a $Z^\prime$ when kinematics allows it. In particular, we consider photons from neutral meson decays and electron bremsstrahlung. On the other hand, positrons can lead to resonant production of $Z^\prime$ through on-shell annihilation with electrons. In general, We find that the production channel from meson decays is almost negligible, but it provides a useful term of comparison with the results available in literature. Here we do not consider $Z^\prime$ production through Compton scattering \cite{Celentano:2020vtu}, since it does not change our conclusions in the relevant region of the parameter space. We also neglect charged meson decays, but in Section V we compare our final results with those obtained in \cite{Berryman:2019dme} where such a channel is studied in detail.

In this Section, we provide a brief description of the calculation method we employ for each channel. First, let us define $N_{\gamma}^{a,ij}$ as the number of photons in the $i$-th energy bin and $j$-th angular bin, from the production channel $a$, as predicted by GEANT4, where the angle is formed by the original proton beam and the outgoing photon propagation directions. We also define the bin extrema to be $[E^{\rm min}_i,E^{\rm max}_i]$ and $[\theta_j^{\rm min},\theta_j^{\rm max}]$ for $i$-th energy and $j$-th angular bins, respectively. For meson decays and bremsstrahlung, we assume that the $Z^\prime$ has the same angular and energy distributions of the corresponding photons. Analogously, we define $N_{e^+}^{ij}$ as the number of positrons and assume that all the $Z^\prime$ from resonant production have the same propagation direction of the incoming positron.

\subsection{Neutral Meson Decay} 
The number of $Z^\prime$ in the $ij$-th bin produced by neutral meson decays can be estimated by
 \begin{equation}
 N_{Z^\prime}^{M,ij}=N_{\gamma}^{M,ij}Br(M\to\gamma Z^\prime)\,,
 \label{N_meson_bin}
 \end{equation}
where $M=\pi^0,\eta^0$ and $Br(M\to\gamma Z^\prime)$ is the branching ratio of the neutral meson decaying into a photon and a $Z^\prime$, given by
 \begin{equation}
 Br(M\to\gamma Z^\prime)=2\epsilon^2(m^2_{Z^\prime})\left(1-\frac{m^2_{Z^\prime}}{m^2_{M}}\right)^3 Br(M\to\gamma \gamma)\,,
 \label{branching_ratio_meson}
 \end{equation}
 where $\epsilon$ is calculated using Eq. \ref{kinetic_mixing}.
 We take $Br(\pi^0\to\gamma \gamma)=0.98823$ and $Br(\eta^0\to\gamma \gamma)=0.3941$ \cite{Zyla:2020zbs}. 

\subsection{Bremsstrahlung}
To calculate the number of $Z^\prime$ in the $ij$-th bin produced by electron and positron bremsstrahlung, we adopt the same approach proposed in \cite{Dutta:2020vop}
 \begin{equation}
 N_{Z^\prime}^{\text{brem},ij}=N_{\gamma}^{\text{brem},ij}\left(\frac{g}{e}\right)^2f\left(\frac{m_{Z^\prime}}{\langle E_e\rangle}\right)\,,
 \label{N_brem_bin}
 \end{equation}
 where $e$ is the electric charge, the function $f(x)=1154\exp(-24.42 x^{0.3174})$ is taken from Fig. 9 in \cite{Dutta:2020vop} and represents a phase space factor, and $\langle E_e \rangle=1.0773E_\gamma + 13.716\,[\text{MeV}]$ is the average electron or positron energy. $g$ is the coupling strength to electron and positrons, which depends on the model under consideration: $g=g_{\mu e}\,(g_{e\tau})$ for $L_\mu-L_e$ ($L_e-L_\tau$), $g=e\,\epsilon(m^2_{Z^\prime})$ for $L_\mu-L_\tau$. 

\subsection{Resonant Production}

\begin{figure*}[]
   \includegraphics[width=\columnwidth]{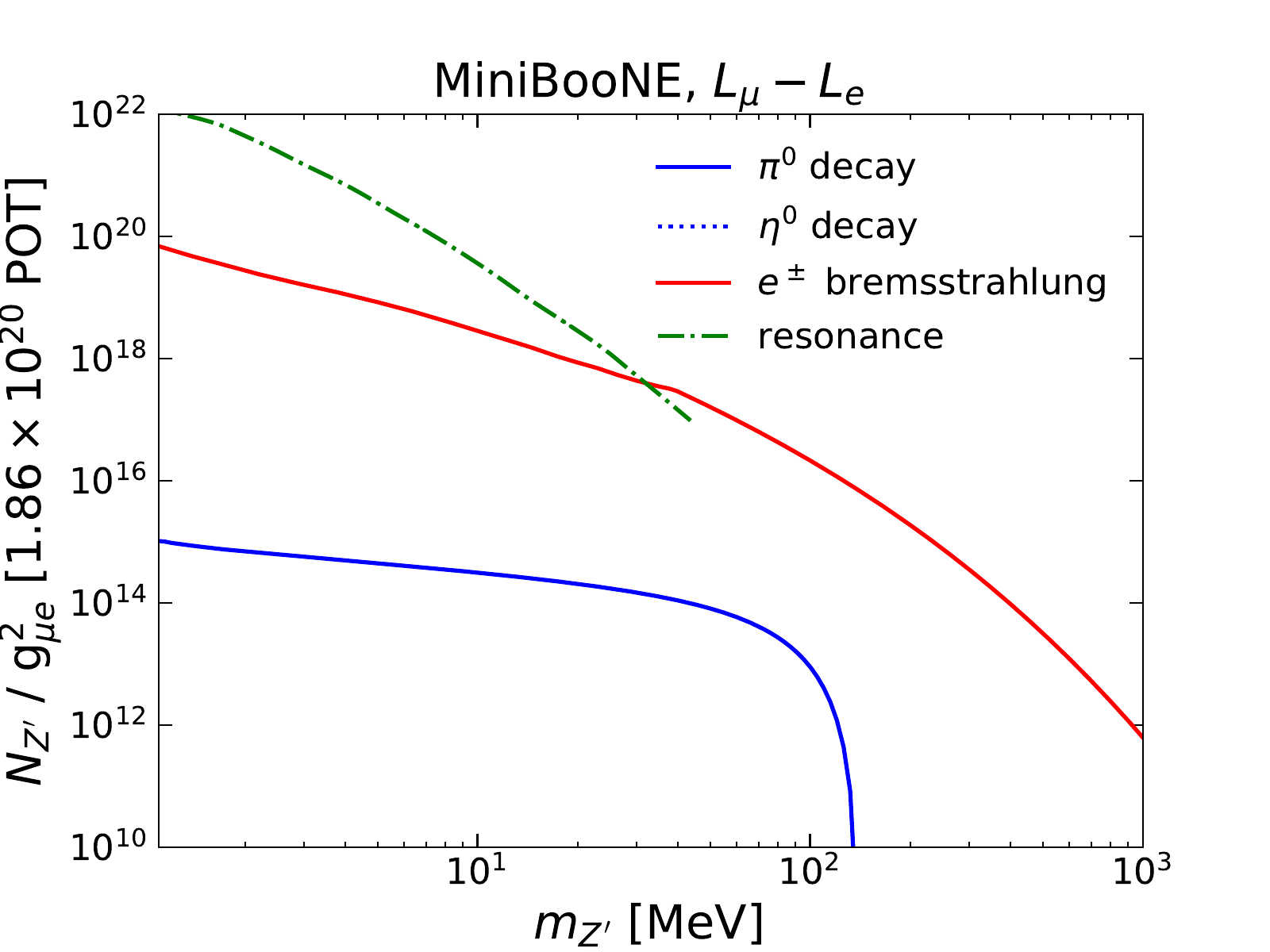}
   \includegraphics[width=\columnwidth]{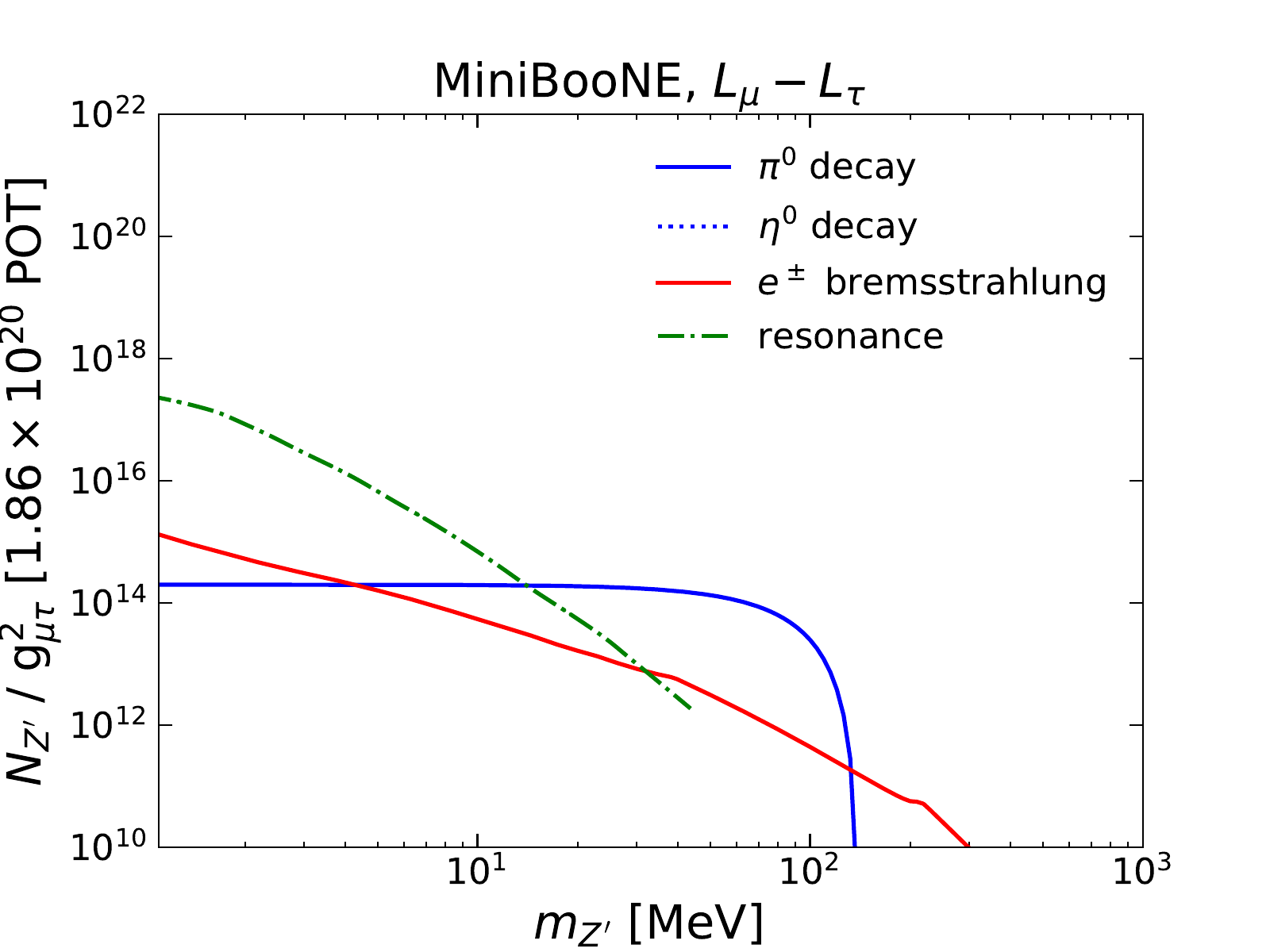}
   \includegraphics[width=\columnwidth]{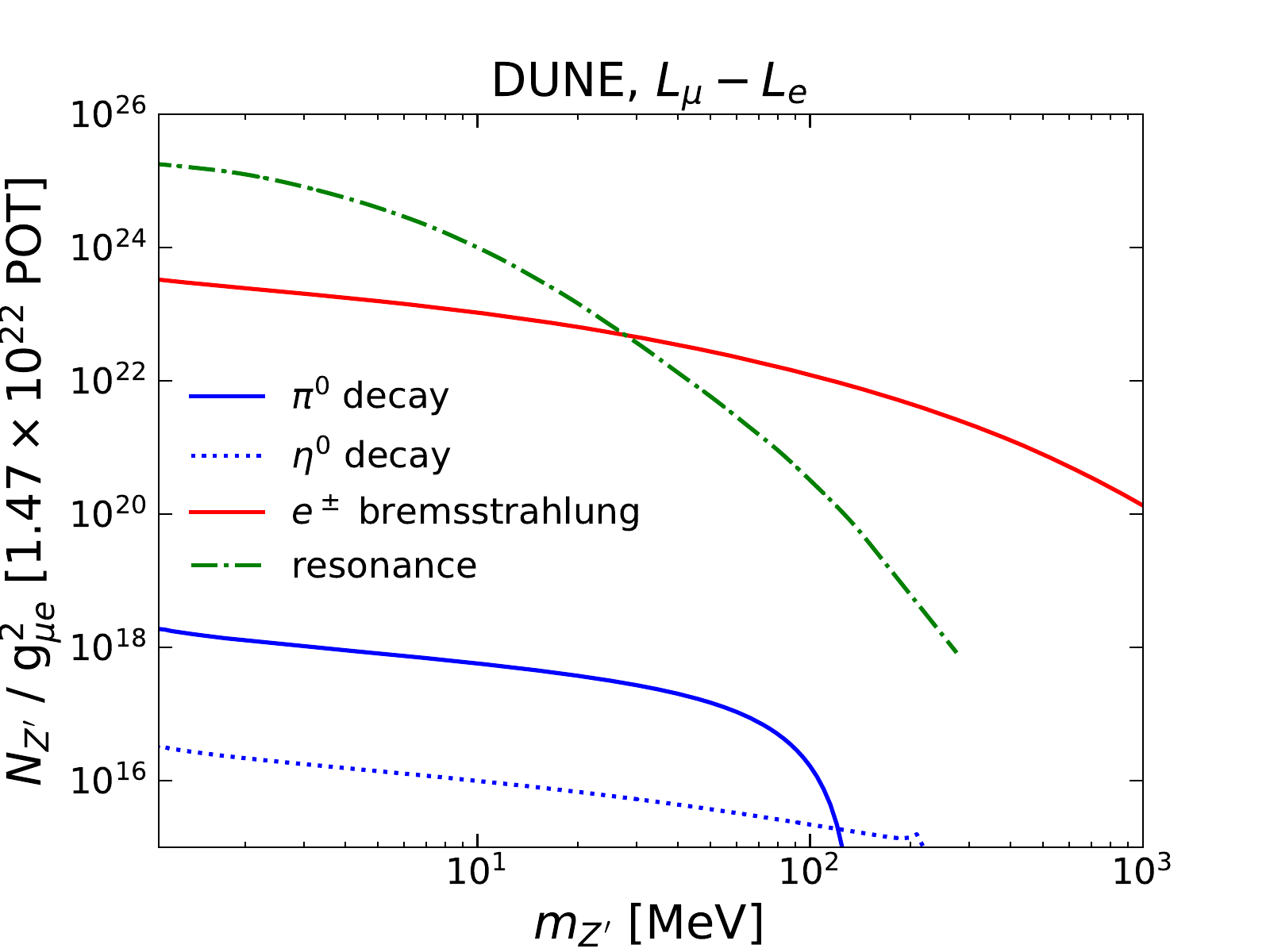}
   \includegraphics[width=\columnwidth]{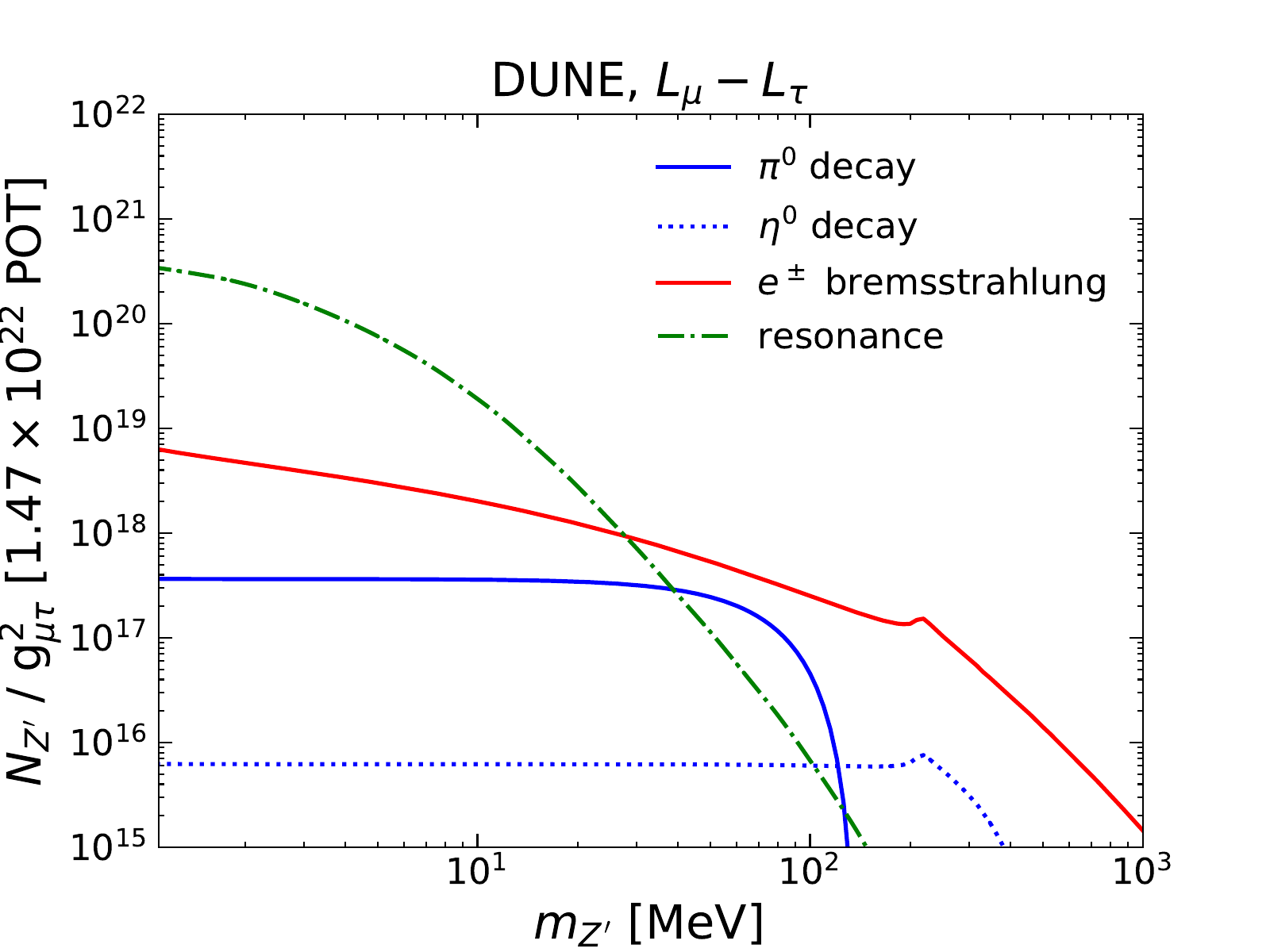}
  \caption{(Top) Number of $Z^\prime$ entering-jy the MiniBooNE
 detector, as a function of $m_{Z^\prime}$ for each production channel. Left panel refers to $L_\mu-L_e$, whereas the right one refers to $L_\mu-L_\tau$. (Bottom) Same as the panels in the top row, but for DUNE.}
  \label{fig:Zprime_number_comparison}
\end{figure*}

A $Z^\prime$ can be produced on-shell through the process $e^+ + e^-\to Z^\prime$ when $E^{\rm res}_{e^+}=E_{Z^\prime}^{\rm res}=m_{Z^\prime}^2/2m_e$. In this case the number of $Z^\prime$ in the $j$-th angular bin is given by
 \begin{equation}
 N_{Z^\prime}^{\text{res},j}=\frac{AX_0}{m_pZ}\sum_{i}\int_{0}^{t_{\rm max}} dtN_{e^+}^{ij}I(E_{i},E_{e^+}^{\rm res},t)\sigma_{\rm res}\,,
 \label{N_resonance_bin}
 \end{equation}
where $A$ and $Z$ are the mass and atomic number of the nuclei in the proton beam target (or beam dump), respectively, $X_0$ is the radiation length of the same target, $m_p$ is the mass of proton, $I(E_{i},E_{e^+},t)$ is the probability that a positron with initial energy $E_i$ (the average energy of the $i$-th bin) has a final energy $E_{e^+}$ after propagating $t$ radiation lengths, and $t_{\rm max}$ is the maximum number of radiation lengths traveled by a positron in the target. The latter probability is taken from Eq.~\ref{eq:current} in \cite{Nardi:2018cxi}. $\sigma_{\rm res}$ is the cross section for resonant production and is given in \cite{Celentano:2020vtu}
\begin{equation}
\sigma_{\rm res}=\frac{\pi g^2}{2m_e}\delta\left(E_{e^+}-\frac{m^2_{Z^\prime}}{2m_e}\right)\,,
\label{sigma_res}
\end{equation}
where $g=g_{\mu e}\,(g_{e\tau})$ for $L_\mu-L_e$ $(L_e-L_\tau)$, $g=e\,\epsilon(m^2_{Z^\prime})$ for $L_\mu-L_\tau$. 

For the sake of simplicity, in the case of DUNE, we use a single value for $t_{\rm max}$, regardless of the original production point of positrons in the target. This average ($t_{\rm max}$) is calculated as the radiation lengths of the target downstream of the 50\% positron production point.  The final result is $t_{\rm max} = 3.3$. In the case of MiniBooNE, considering that secondary particles are propagating through the relatively thick beam dump, we take $t_{\rm max} = 5$. Larger values of $t_{\rm max} = 5$ would not lead to any significant difference. 
 
 \begin{figure}[b]
  \includegraphics[width=\columnwidth]{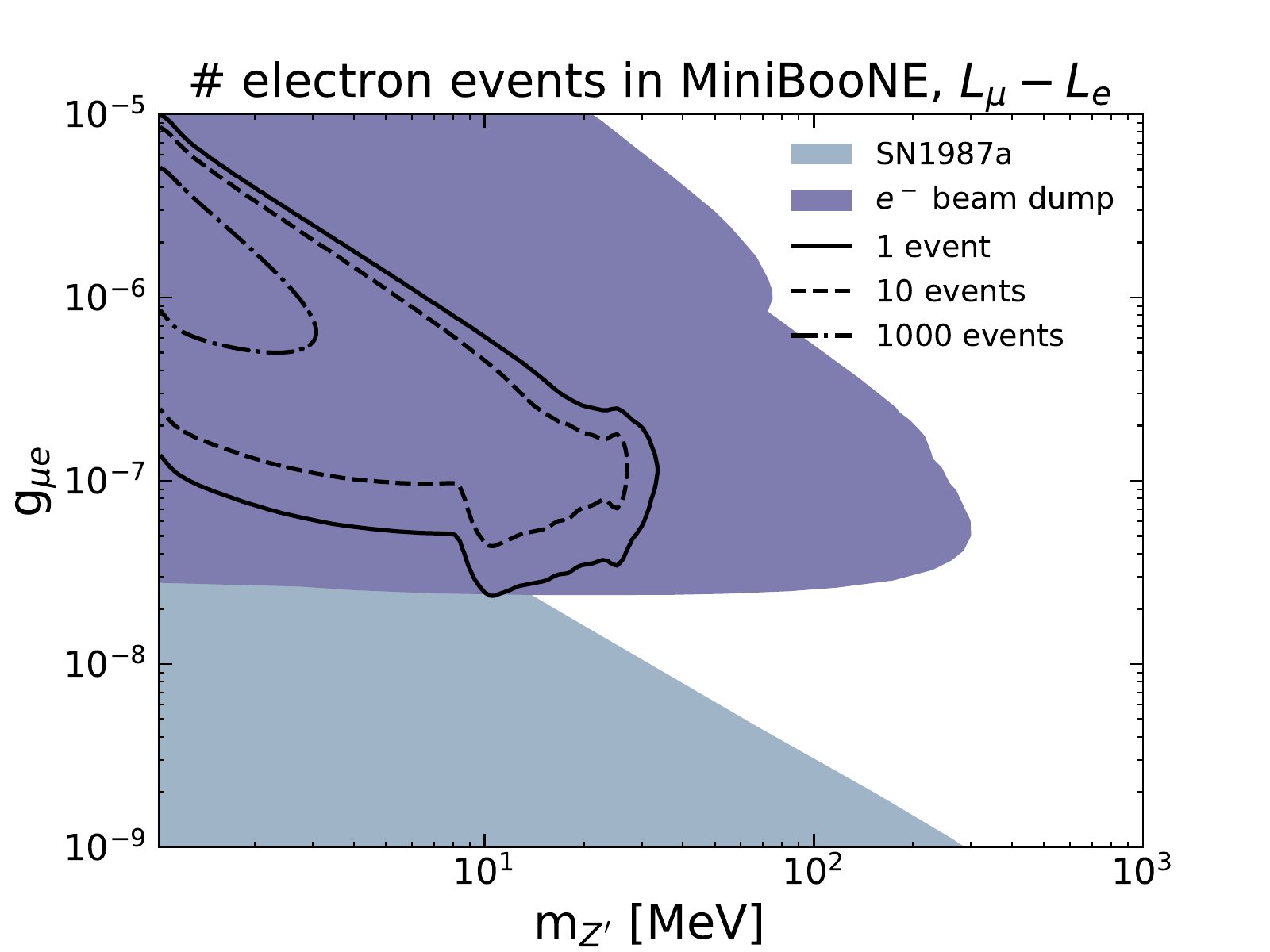}
  \caption{The black contours display the number of electron events in the MiniBooNE detector from $Z^\prime$ decays as a function of $m_{Z^\prime}$ and the coupling constant $g_{\mu e}$ in the context of $L_\mu-L_e$.  The solid filled areas represent the region of the parameter space already excluded by electron beam dump experiments \cite{Bauer:2018onh} and SN1987a \cite{Croon:2020lrf}. See section VI for more details concerning such external constraints.}
  \label{fig:MiniBoone_Lmu_Le_electrons}
\end{figure}

\begin{figure*}[]
  \includegraphics[width=\columnwidth]{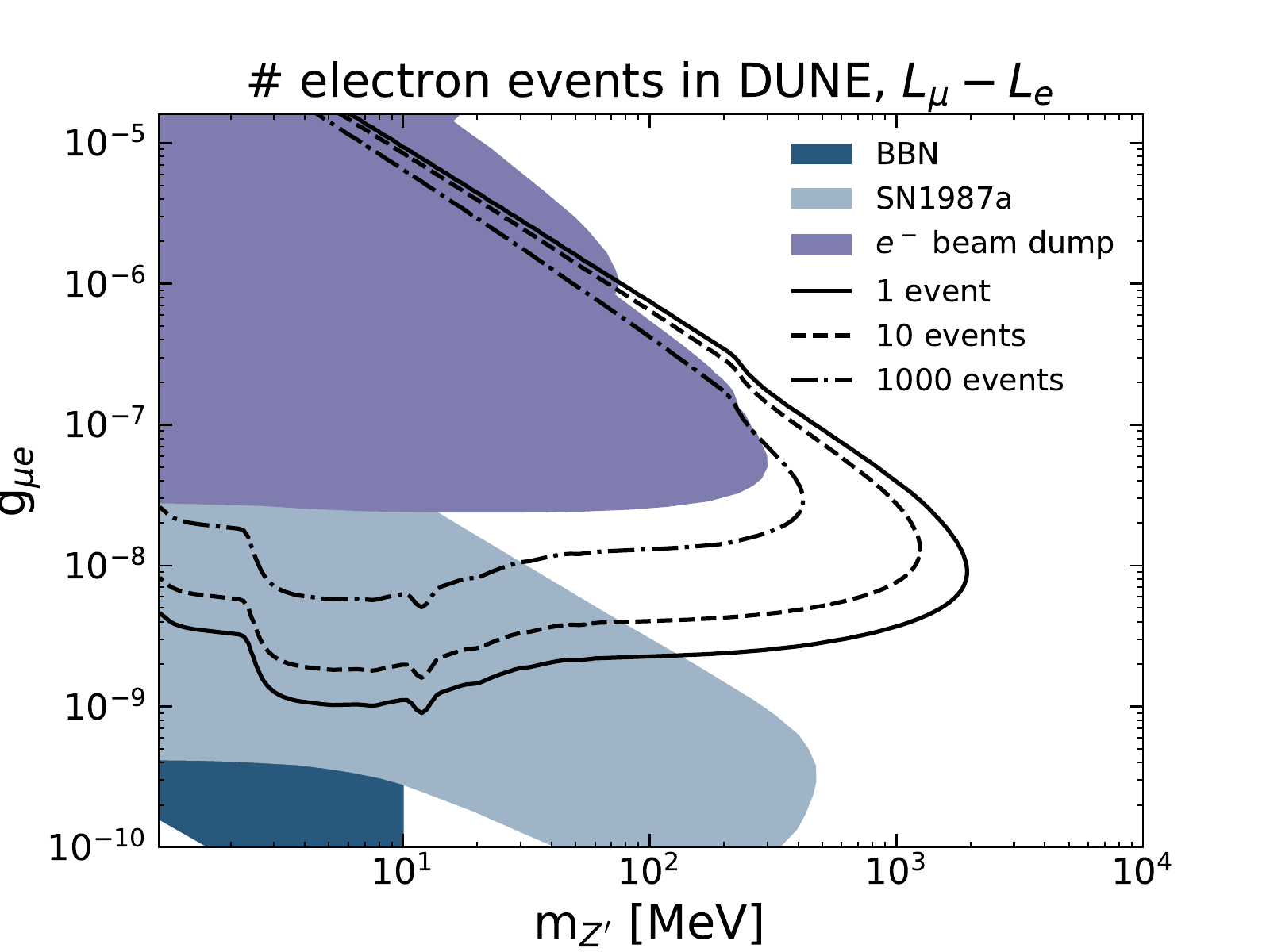}
  \includegraphics[width=\columnwidth]{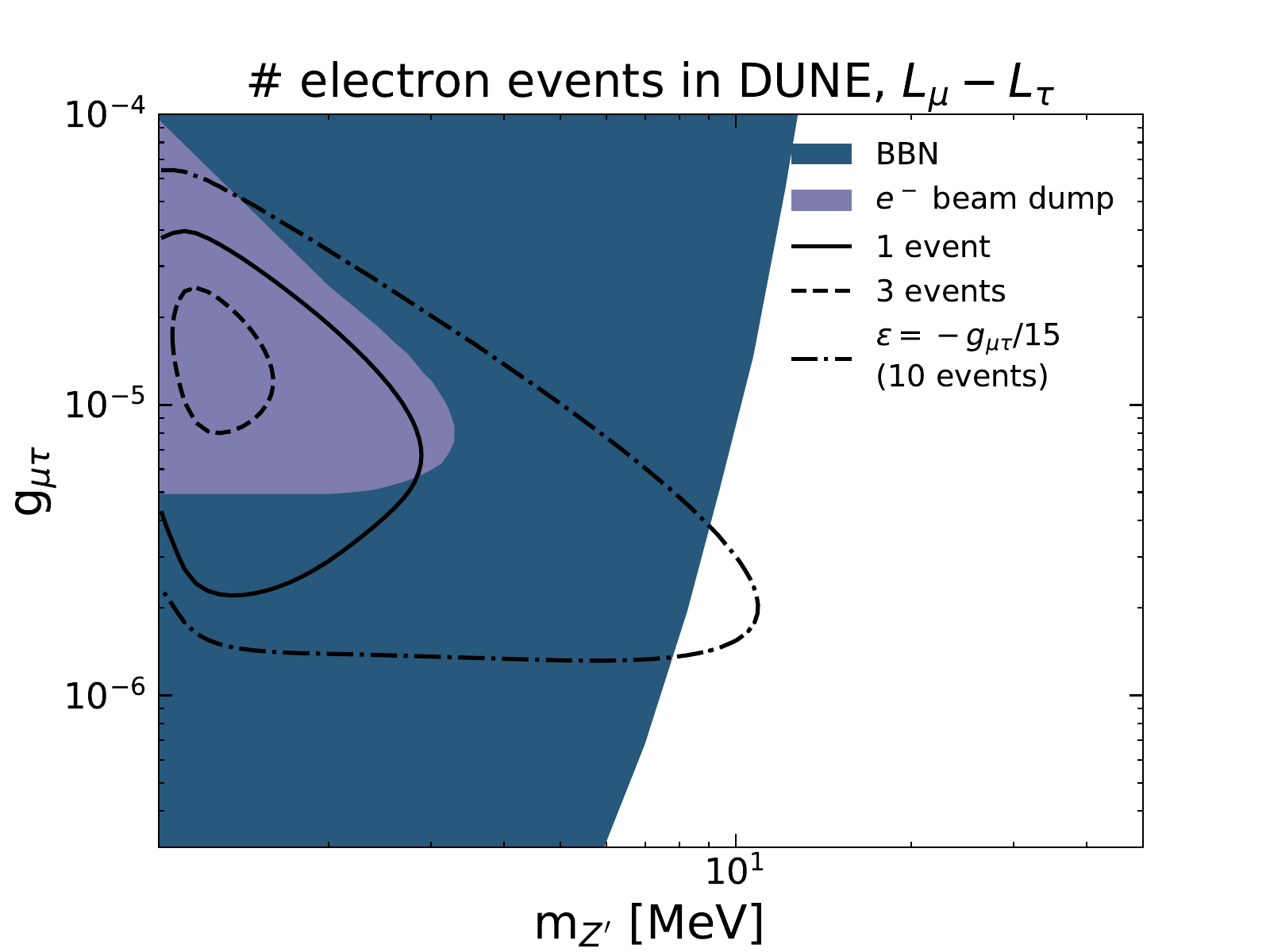}
  \caption{(Left) Same as Fig. \ref{fig:MiniBoone_Lmu_Le_electrons}, but for DUNE. The BBN bounds are taken from \cite{Knapen:2017xzo}. (Right) Same as the left panel, but for $L_\mu-L_\tau$. In this case, the BBN constraints are taken from \cite{Escudero:2019gzq}.}
  \label{fig:DUNE_Lmu_Le_electrons}
\end{figure*}

\subsection{Comparison of production channels}

The top left panel of Fig. \ref{fig:Zprime_number_comparison} displays the number of $Z^\prime$ as a function of $m_{Z^\prime}$ in the context of $L_\mu-L_e$ from each production channel, entering the MiniBooNE detector. The number of $Z^\prime$ is in units of $g^{2}_{\mu e}$ and refers to an exposure of $1.86\times 10^{20}$ protons on target (POT). For $m_{Z^\prime}< 20$ MeV, the dominant channel is resonant production, whereas bremstrahlung provides the biggest contribution for higher masses. Resonant production decreases faster since only positrons with energy $E_{e^+}^{\rm res}=m_{Z^\prime}^2/(2m_e)$ are able to produce the $Z^\prime$ on shell, whereas for bremsstrahlung all charged leptons with $E_{e^\pm}>m_{Z^\prime}$ can in principle contribute. Neutral meson decays are subdominant because they occur only through kinetic mixing, and their number per each proton on target is smaller than the one of electrons and positrons.

The top right panel of Fig. \ref{fig:Zprime_number_comparison} refers to $L_\mu - L_\tau$. In this case, the pion decay channel is not significantly modified with respect to $L_\mu-L_e$. On the other hand, both bremsstrahlung and resonant production take place only through kinetic mixing and they receive a $\sim 10^4$ suppression factor.

The bottom panels of Fig.~\ref{fig:Zprime_number_comparison} refer to DUNE. In this case, the number of $Z^\prime$ is normalized to $1.47\times 10^{22}$ POT. Apart from a higher number of POT, DUNE has a 120 GeV proton beam, whereas MiniBooNE uses one with 8 GeV, therefore in the former case, we have an overall increased number of $Z^\prime$, especially for large masses which are kinematically suppressed in MiniBooNE. Finally, we point out that for $L_e-L_\tau$ (not shown), we expect the same number of  $Z^\prime$ obtained for $L_\mu-L_e$.

\section{$Z^\prime$ decays in the detector}
\label{sec:detector}
The number of leptons ($l=e,\mu$) produced by $Z^\prime\to l^+l^-$ decays in the detector is given by
\begin{equation}
\begin{split}
N_{l}=\sum_{j}N_{Z^\prime}^{\text{res},j}n_{l}P_{Z^\prime\to l^+l^-}(E_{Z^\prime}^{\rm res},\theta_j)\\
+\sum_{i,j}n_{l} \left(N_{Z^\prime}^{\text{brem},j} + N_{Z^\prime}^{\pi^0,\eta^0,j}\right) P_{Z^\prime\to l^+l^-}(E_{Z^\prime}^i,\theta_j)
\end{split}
\label{number_of_leptons_from_decay}
\end{equation}
where $n_{l}=n_l(E_{Z^\prime},E_l^{\rm th},\theta_{1},\theta_{2})$ is the number of leptons per $Z^\prime$ decay with an energy $E_l$ greater than the detection threshold $E_l^{\rm th}$ and going into an angle cone between $\theta_{1}$ and $\theta_{2}$ (see Eq. \ref{number_of_leptons_per_decay}),  $E_{Z^\prime}^{\rm res}$ is the energy of the $Z^\prime$ for resonance production, $E^i_{Z^\prime}=\frac{1}{2}(E_i^{\rm min}+E_i^{\rm max})$ is the center of the $i$-th energy bin, and $P_{Z^\prime\to l^+l^-}$ is the probability that a $Z^\prime$ decays inside the detector. The latter is calculated with the following equation
\begin{equation}
\begin{split}
P_{Z^\prime\to l^+l^-}(E_{Z^\prime},\theta)=\left(1 - e^{-\frac{L(\theta) \Gamma(Z^\prime\to l^+l^-)m_{Z^\prime}}{p_{Z^\prime}}}\right)\\
e^{-\frac{d(\theta)\, \Gamma(Z^\prime\to l^+l^-)m_{Z^\prime}}{p_{Z^\prime}}}\frac{\Gamma(Z^\prime\to l^+l^-)}{\Gamma_{\rm tot}}\,,
\end{split}
\label{decay_probability}
\end{equation}
where $L(\theta)$ is distance traveled in the detector, which depends on the $Z^\prime$ propagation angle $\theta$, $d(\theta)$ is the distance traveled between the target (or beam dump) and the detector, $p_{Z^\prime}$ is the momentum of the $Z^\prime$, $\Gamma_{\rm tot}$ and $\Gamma(Z^\prime\to l^+l^-)$ are the total and partial decay widths, respectively. 

In Eq. \ref{number_of_leptons_from_decay}, the sum over $j$ is performed considering only those propagation directions of the $Z^\prime$ within the detector coverage, i.e. those having a propagation angle with respect to the beam $\theta_{Z^\prime}<\theta_{\rm det}$. Assuming the symmetry axis of a detector is aligned with the direction of the beam and that it has a width $2w$, we can estimate its angular size as $\theta_{\rm det}\sim \frac{w}{d}$.

\begin{table*}[t]
\caption{\label{Tab:Exp_details} 
\footnotesize  Experimental details of MiniBooNE and DUNE used for evaluating Eq. \ref{number_of_leptons_from_decay} and \ref{decay_probability}. $L$ is the length of the detector, $d$ is the distance from the target or the beam dump to the detector, $w$ is half the width of the detector perpendicular to the proton beam direction, $\theta_{\rm det}$ is the angular size of the detector as seen from the target or the beam dump, $E_l^{\rm th}$ and $E_l^{\rm max}$ are the minimum and maximum energies we consider for charged leptons $l^\pm$ ($l=e,\mu$) from $Z^\prime$ decays, $\theta_1$ and $\theta_2$ are minimum and maximum propagation angles of the the charged leptons, with respect to the proton beam direction, and POT is the number of protons on target.}
\vspace*{0mm}
\centering
\begin{ruledtabular}
\begin{tabular}{c|ccccccccc}
Experiment & $L$ [m] & $d$ [m] & $w$ [m] & $\theta_{\rm det}$ [rad] & $E_l^{\rm th}$ [MeV] & $E_l^{\rm max}$ [MeV] & $[\theta_1,\theta_2]$ [rad] & Efficiency & POT\\
\hline
MiniBooNE \cite{Aguilar-Arevalo:2018wea}& 12 & 490 & 6 & 0.012 & 75 & 850 & [0,0.14] & 10\% & $1.86\times 10^{20}$\\
DUNE \cite{Berryman:2019dme}& 5 & 579 & 2.5 & 0.004 & 0.2 ($e^\pm$), 2 ($\mu^\pm$) & - & [0,$\frac{\pi}{36}$] & 100\% &  $1.47\times 10^{22}$\\
\end{tabular}
\end{ruledtabular}
\end{table*}

The partial decay width to charged leptons is 
\begin{equation}
\Gamma(Z^\prime\to l^+l^-) = \frac{g_{l}^2 m_{Z^\prime}}{12\pi} \sqrt{1 - 4\left(\frac{m_{l}}{m_{Z^\prime}}\right)^2} \left[1 + 2 \left(\frac{m_{l}}{m_{Z^\prime}}\right)^2\right]\,,
\label{partial_decay_width_leptons}
\end{equation}
where $l=e,\mu,\tau\,$, $g_{e}=g_{\mu}=g_{\mu e}$ for $L_\mu-L_e$, whereas $g_{e}=e\,\epsilon(m^2_{Z^\prime})$ and $g_{\mu}=g_{\mu\tau}$ for $L_\mu-L_\tau$. For a neutrino $\nu_l$, the partial decay width is equal to the one in Eq.~\ref{partial_decay_width_leptons} divided by 2. For both models, the total decay width is given by the sum of the partial widths of each lepton, since the contribution from hadrons is negligible.

    \section{MiniBooNE and DUNE sensitivities}
\label{sec:sens}

Let us first consider MiniBooNE. We consider the data sample corresponding to $1.86\times 10^{20}$ POT collected by the MiniBooNE collaboration in the beam dump mode \cite{Aguilar-Arevalo:2018wea}. We focus on the data corresponding to the $\nu-e$ elastic scattering channel which contains only 2 events, as reported in Fig. 18 of Ref. \cite{Aguilar-Arevalo:2018wea}. This sample refers to electrons with energies between 75 and 850 MeV, which are produced in a narrow angular cone $\cos\theta>0.99$ around the original proton beam direction. Adopting such kinematic cuts, a 10\% efficiency in $e^\pm$ detection and other experimental properties listed in Table \ref{Tab:Exp_details}, we evaluate Eq. \ref{number_of_leptons_from_decay} as a function of $m_{Z^\prime}$ and of the coupling. 

Figure \ref{fig:MiniBoone_Lmu_Le_electrons} shows the results of such a calculation in the case of $L_\mu-L_e$ in black contours. In general, all the electrons come from $Z^\prime$ produced through bremsstrahlung, except for $m_{Z^\prime}\gtrsim 9$ MeV and $g_{\mu e}<10^{-7}$ in which the dominant contribution comes from resonant production. The reason why the latter plays a role only for $m_{Z^\prime}\gtrsim 9$ MeV is that to produce electrons with energy $E_l>E_{l}^{\rm th}$ one needs a $E^{\rm res}_{Z^\prime}=m^2_{Z^\prime}/(2m_e)$ greater than the energy threshold of 75 MeV. The solid colored regions represent the area of the parameter space ($m_{Z^\prime},g_{\mu e}$) already excluded. We give more details about such constraints in Section\ref{sec:cons}. 

A remark is in order. In Fig. \ref{fig:Zprime_number_comparison}, we show that resonant production channel dominates    for $m_{Z^\prime}\in [0,30]$ MeV. This seems in contrast with what displayed in Fig. \ref{fig:MiniBoone_Lmu_Le_electrons}, where this channel only contributes from $m_{Z^\prime}\gtrsim 9$ MeV. This apparent disagreement is justified by the fact that in Fig. \ref{fig:Zprime_number_comparison} we show only the number of $Z^\prime$ that are travelling in the solid angle identified by the detector, without taking into account any decay and kinematic cuts to the decay products.

Assuming to be in the currently allowed region of the parameter space, $Z^\prime$ decays in MiniBooNE provide no more than 1 electron event, which is comparable to the $\sim 2$ events expected as background. Therefore, the bounds one can derive from MiniBooNE are similar to current ones for $L_\mu-L_e$. For $L_\mu-L_\tau$ (not shown), for which both the $Z^\prime$ production and decay channels involving electrons are suppressed by kinetic mixing, no constraint can be set. In both models, the expected number $\mu^\pm$ produced is negligible.

For DUNE, we assume an exposure of $1.47\times 10^{22}$ POT and study $Z^\prime$ decays in the multipurpose near detector filled with gaseous argon (GAr). Concerning detection thresholds, we take $0.2$ MeV and $2$ MeV for $e^\pm$ and $\mu^\pm$, respectively \cite{Berryman:2019dme}, which stem from the assumption of 2 cm as a reasonable length for track identification. These and other experimental specifications used in evaluating Eq. \ref{number_of_leptons_from_decay} are summarized in Table \ref{Tab:Exp_details}. A careful evaluation of backgrounds in the GAr detector is mandatory. This has already been done in \cite{Berryman:2019dme}. 
Here we just report a brief summary. Concerning $Z^\prime$ decays to $e^\pm$, the main background comes from neutral pions produced in neutrino interactions, for which few $\times 10^5$ events per ton-year are expected. 

Despite the large statistics, a $\pi^0$ event can be misinterpreted only under the following circumstances. The first possibility is the absence of other hadronic activity, or one photon from $\pi^0$ is either missed by the calorimeter surrounding the GAr detector or cannot be associated to the same vertex of the other photon. 
The second case occurs when only one of the two photons convert in the gas, which occurs 12\% of the time. Taking into account such requirements, the number of background events is reduced to few thousands. Further reduction might come from considering only those events inside an angular cone centered on the original proton beam direction, since the nuclear processes producing neutral pions are usually isotropic, whereas the $Z^\prime$ decay is mostly forward. If one considers the opening angle of this cone to be equal to the angular resolution (4 mrad), then the number of photons from $\pi^0$ is suppressed to 0.01\%. Even a conservative 5$^\circ$ angular cut provides about 50 background events. Here, we propose to use 10 $e^\pm$ events as nominal threshold for defining the constraints on leptophilic gauge bosons.

For $Z^\prime$ decaying to $\mu^\pm$, the main background comes from muon neutrinos charged current (CC) interactions with pion production. For these events, both the presence of nuclear activity and kinematic cuts can significantly reduce the number of background events, which is otherwise $O(10^5)$ per ton-year. Furthermore, the calorimeter surrounding the argon gas will have the capability to distinguish between muon and pion events, with a probability of 70\% \cite{Berryman:2019dme}. Finally, adding the proposed muon tagger after the GAr detector would strongly improve the rejection of background. In the end, we adopt the same 10 events $\mu^\pm$ threshold, as done for $e^\pm$ pairs and as in \cite{Berryman:2019dme} to be conservative.

The expected number of electrons from $Z^\prime$ decays in DUNE for $L_\mu-L_e$ and $L_\mu-L_\tau$ is shown in the left and right panel of Fig. \ref{fig:DUNE_Lmu_Le_electrons}, respectively. In contrast to MiniBooNE, for $L_\mu-L_e$ the contribution from resonant production dominates over bremsstrahlung for $m_{Z^\prime}>2$ MeV. The reason is that  in the resonance case the $Z^\prime$ has a specific energy ($m^2_{Z^\prime}/2m_e$), which for $m_{Z^\prime}<2$ MeV is not enough to allow for a decay to $e^\pm$ in a 5 degree cone (see the $\theta_1,\theta_2$ column of Table \ref{Tab:Exp_details}).  According to our nominal 10 event threshold, DUNE can extends beyond the current exclusion region from electron beam dump experiments to values of $g_{\mu e}$ ten times smaller in the mass range $m_{Z^\prime}\in[1,10^3]$ MeV. While part of the parameter space is disfavored by observations of neutrinos from SN1987a \cite{Croon:2020lrf}, it is worthwhile to note that the DUNE sensitivity extends up to significantly higher $m_{Z^\prime}$. 

In the $L_\mu-L_\tau$ case, on the other hand, the improvement over the current constraints dominated by electron beam dump experiments is marginal. Nevertheless, as shown in \cite{Berryman:2019dme}, the $Z^\prime$ production channels from charged meson decays to muons ($M^\pm\to\mu^\pm\nu_\mu Z^\prime$, $M=\pi,K$) are not suppressed by kinetic mixing and a more significant improvement can be obtained. This hierarchy of production channels can change if one allows for a tree level contribution to the kinetic mixing. For instance, in the right panel of Fig. \ref{fig:DUNE_Lmu_Le_electrons}, we show that for $\epsilon=-g_{\mu\tau}/15$ (dot-dashed line) the sensitivity of DUNE would be comparable to the BBN constraints \cite{Escudero:2019gzq}. In this case, however, the electron beam dump constraints also need to be properly rescaled.

\begin{figure}[h]
  \includegraphics[width=\columnwidth]{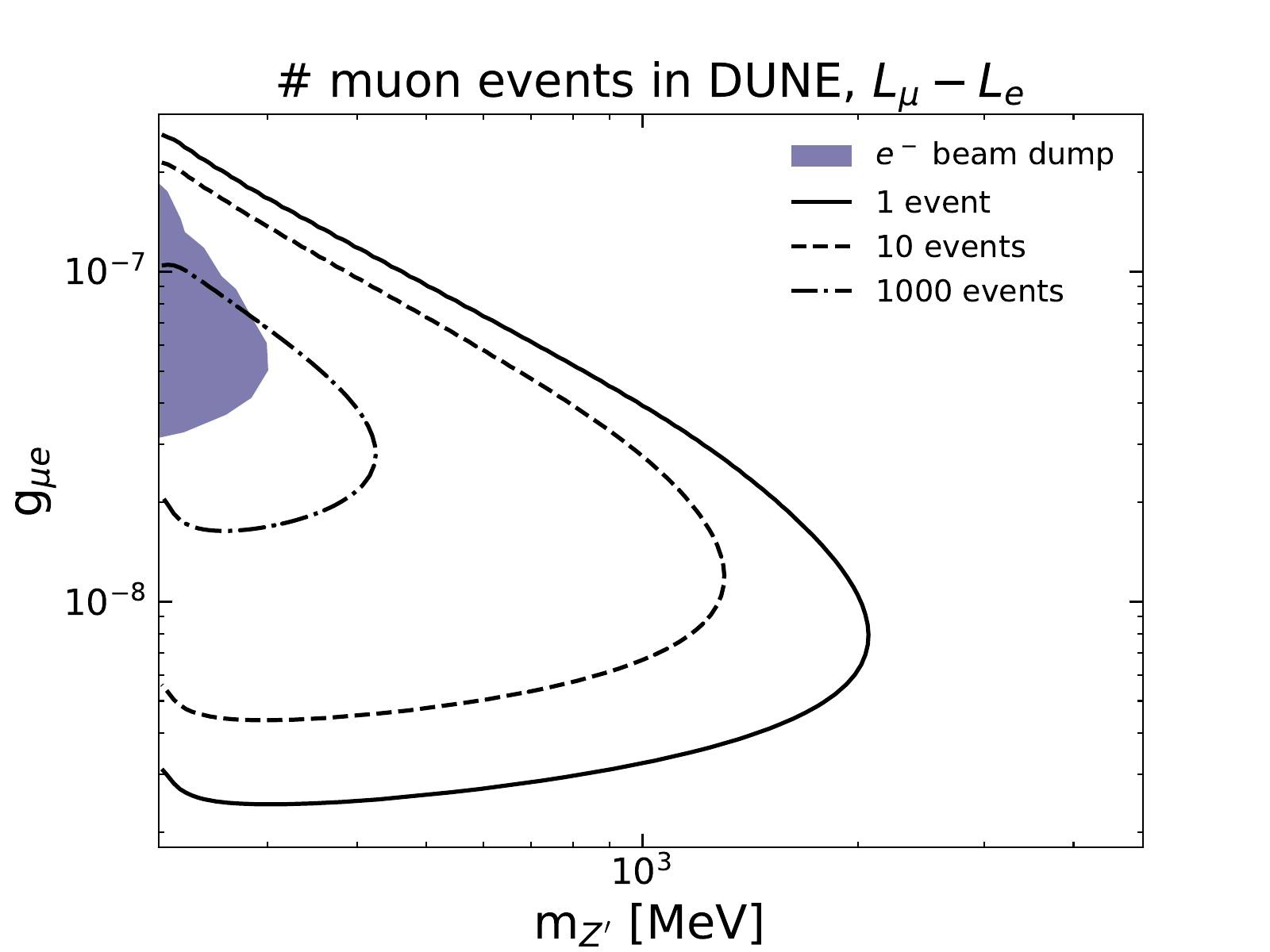}
  \caption{Same as the left panel of Fig. \ref{fig:DUNE_Lmu_Le_electrons} but for the muons produced by $Z^\prime$ decaying in the detector.}
  \label{fig:Lmu_Le_muons}
\end{figure}

Figure \ref{fig:Lmu_Le_muons} displays the number of muons produced in the $L_\mu-L_e$ case. This number is comparable to the one of electrons shown in the left panel of Fig. \ref{fig:DUNE_Lmu_Le_electrons} for the mass range $m_{Z^\prime}>200$ MeV. This is expected because the branching ratios to electrons and muons are equal to each other. 

In our sensitivity study of DUNE, so far we have not yet discussed  
the $Z^\prime$ decays in the liquid argon near detector (LAr). Such a detector can in principle observe the number of decays comparable to the one for the GAr detector, due to its similar volume and distance from the production target. The LAr detector, however, will suffer from larger backgrounds resulting from higher number of neutrino interactions than in the GAr detector. Nevertheless, since the LAr detector will be the first available for data taking for DUNE, it makes sense to provide a brief discussion concerning its potential. As for the GAr detector, the main background for $e^\pm$ pairs is $\pi^0$ decay. Assuming $\pi^0$ can be reconstructed with a high degree of accuracy as in MicroBooNE~\cite{MicroBooNE:2019rgx}, one can remove them efficiently using suitable invariant mass choices. Consequently, a large region of the parameter space ($m_{Z^\prime}> 200$ MeV) can still be explored with a similar sensitivity to that of the GAr, because of the similar number of observable $Z^\prime$ decays.

\section{Main results}\label{sec:key}
Prospective constraints of DUNE on both $L_\mu-L_e$ and $L_\mu-L_\tau$ were already derived in \cite{Berryman:2019dme}, thus a comparison with the results obtained therein is in order. The main difference our study, however, lies in the production channels of $Z^\prime$ under consideration. In our case, the main channels are the electron bremsstrahlung and the resonant production. These channels have not been considered in \cite{Berryman:2019dme} where instead the bulk of $Z^\prime$ comes from charged meson decays. For $L_\mu-L_e$ we find that the number of $Z^\prime$ produced in the target is five orders of magnitude larger than the one connected to charged mesons estimated in \cite{Berryman:2019dme}. This is clearly seen by comparing our bottom left panel of Fig. \ref{fig:Zprime_number_comparison} with Fig. 5.2 in \cite{Berryman:2019dme}. Such a difference explains why we conclude that DUNE can improve current constraints in the parameter space ($m_{Z^\prime},g_{\mu e}$), whereas the opposite is stated in \cite{Berryman:2019dme}. Comparing our calculation for $L_\mu-L_\tau$, shown in the bottom right panel of Fig. \ref{fig:Zprime_number_comparison}, with Fig. 5.4 in \cite{Berryman:2019dme}, we find that the number of $Z^\prime$ from charged mesons is similar to the one from electrons. This happens because the charged meson decays involving muons are not suppressed by kinetic mixing, whereas a suppression is indeed present for both the bremsstrahlung and resonant channels.

A remark is in order. Both here and in \cite{Berryman:2019dme}, the number of $Z^\prime$ produced by neutral meson decays is calculated, but we find it to be about one order of magnitude larger.  A similar comparison can be made with Fig. 6 in \cite{Celentano:2020vtu}, though in the context of a generic kinetic mixing model and only for MiniBooNE. In this case, the difference is just a factor of two. The reason for the larger discrepancy with \cite{Berryman:2019dme} is probably due to a different numerical approach: we have used GEANT4 as a tool to simulate secondary particle production in proton interactions, whereas in \cite{Berryman:2019dme} Pythia is employed. As far as we understand, Pythia is harder to use for our purposes, because it simulates only a single event of particle collisions, thus essential inputs like target geometry or target material are not required. We emphasize that GEANT4 is also used in \cite{Celentano:2020vtu} with which we are in relatively good agreement.

Finally, despite not being explicitly displayed, the sensitivities of DUNE and MiniBooNE in the context of $L_e-L_\tau$ are expected to be the same as $L_\mu-L_e$. This happens because the dominant production channels (bremsstrahlung and resonance) are characterized by the same coupling strength $g_{e\tau}=g_{\mu e}$. 

\section{Current constraints on $L_{\alpha} - L_\beta$}
\label{sec:cons}
Currently, in our region of interest the best sensitivity is obtained in the electron beam dump experiments E137 \cite{Bjorken:1988as} and Orsay \cite{Davier:1989wz}. In both cases, the bounds are taken from \cite{Bauer:2018onh} where a recasting of the original limits for dark photons \cite{Bjorken:2009mm,Andreas:2012mt} has been performed following the approach proposed in \cite{Ilten:2018crw}. 

In terms of astrophysical constraints, lower values of the couplings $g_{\mu e}$ and $g_{\mu\tau}$ are in principle probed by the observations of neutrinos from a core-collapse supernova. The floor of these constraints refer to the case where the extra cooling of the proto-neutron star, induced by $Z^\prime$ production exceeds the one observed in neutrinos. On the other hand, the ceiling of the constraints corresponds to a regime in which $Z^\prime$ bosons are trapped inside the star, and they are re-emitted with a blackbody spectrum from a sphere similar to the neutrinosphere. The observations of about 20 neutrino events from SN1987a has been first used for deriving constraints on $L_\mu-L_\tau$ in \cite{Escudero:2019gzq}. These have very recently been updated in \cite{Croon:2020lrf} taking into account also the non-negligible muon population in the core of the supernova, which allow for an extra source of production and scattering of the $Z^\prime$. 

In this work, we refer to the updated constraints in \cite{Croon:2020lrf} and assume that those for $L_\mu-L_e$ which are not presented therein are equivalent to those obtained in the context of $L_\mu-L_\tau$. This is a conservative assumption for the following reason. Being strongly dependent on the production rate of the $Z^\prime$, the floor of the constraints are not expected to change for $L_\mu-L_e$, because no significant charged $\tau$ leptons are created, and the potential contribution of electrons to the production rate is suppressed by their degeneracy. Electron degeneracy, however, is only present in the core, which means that the $Z^\prime$ scattering rate at larger radii is strongly modified by the direct coupling to $e^\pm$, effectively moving the trapping regime to lower values of the couplings and correspondingly the ceiling of the bounds. An accurate determination of the ceiling, however, would require a dedicated study, which is beyond the scope of this work.

For lower $Z^\prime$ masses ($\leq 10$ MeV), we also have constraints from BBN due to the $Z^\prime$'s decaying into neutrinos and electrons which contribute to both neutrinos and photon evolution at around the MeV scale. The recent accurate measurement of the light degrees of freedom $N_{\rm eff}$ by Planck~\cite{Planck:2018vyg} constrains the coupling and mass of $Z^\prime$, which determine its likelihood to be in the equilibrium  with the SM particles at the MeV scale.   
In our work, for $L_\mu-L_\tau$, we show the BBN limits from Ref.~\cite{Escudero:2019gzq}. In particular, we refer to the bound corresponding to $\Delta N_{\rm eff}>0.2$ which is similar to the current precision of Planck \cite{Planck:2018vyg}. Whereas for $L_\mu-L_e$, we take the constraints derived in Ref~\cite{Knapen:2017xzo} for $B-L$ which also apply to our case.

We emphasize that both the BBN and SN1987a limits are model dependent. For example, the SN1987a can be weakened by the chameleon effect which depends on the environmental matter density~\cite{Nelson:2008tn}, whereas the BBN constraints can be weakened by late reheating~\cite{Depta:2020wmr}.

\section{Conclusions}

The existence of an additional anomaly-free $U(1)$ gauge group which only couple to leptons is a well motivated extension of the Standard Model. Such a model predicts the existence of a new gauge boson $Z^\prime$ which can be dominantly produced by and can decay into leptons. In this work, we have focused on $L_\alpha-L_\beta$, where the new charges are associated to the lepton number difference between flavor $\alpha$ and $\beta$. We have studied current and future constraints from MiniBooNE and DUNE, respectively, on $m_{Z^\prime}$ and $g_{\alpha\beta}$, i.e. the mass and coupling to leptons of the new gauge boson.

For DUNE, we have calculated the number of charged lepton pairs produced by $Z^\prime$ decays in the GAr detector which provides an excellent particle identification and tracking capabilities. We have shown that the $Z^\prime$ produced by $e^\pm$ bremsstrahlung and on-shell resonance give the largest contributions in the context of $L_\mu-L_e$ and $L_e-L_\tau$. This is our main result, since a similar study performed in \cite{Berryman:2019dme} neglected these production channels (focusing mostly on charged mesons decays) and found that DUNE is not able to extend current constraints in the parameter space $[m_{Z^\prime},g_{\mu e}]$ and $[m_{Z^\prime},g_{e\tau}]$. Whereas we find that, when both bremsstrahlung and on-shell resonance are taken into account, DUNE sensitivity goes beyond the current most stringent limits from electron beam dump experiments. We have also shown that such prospective constraints will even extend to larger $m_{Z^\prime}$ beyond the astrophysical bounds from SN1987a and BBN. 

Concerning $L_\mu-L_\tau$, we observe that our $Z^\prime$ production channels endow DUNE with a sensitivity comparable to current constraints. Therefore, in this case the charged meson decay channel presented in \cite{Berryman:2019dme} provides a slightly better reach. These results might change when considering an extra source of kinetic mixing in the context of a UV complete model.

Finally, we have explored the same experimental signatures from $Z^\prime$ decays in MiniBooNE. In particular, we have selected the neutrino-electron elastic scattering data sample collected in the beam dump mode, corresponding to an exposure of $1.86\times 10^{20}$ POT. Despite the very low background (2 events), the maximum number of charged leptons from $Z^\prime$ decays expected in the currently allowed parameter space is $O(1)$. Thus, the sensitivity is comparable to the current bounds.

As future prospects for our work, the following possibilities are available. First, one can study $Z^\prime$ decays in the context of DUNE-PRISM \cite{DUNE:2021tad,DeRomeri:2019kic,Breitbach:2021gvv}, i.e. a near detector complex which will be able to take data at off-axis angles up to a few degrees by moving a set of detectors perpendicular to the beam axis. Since our GEANT4 simulation shows that most of the $Z^\prime$ from bremsstrahlung and resonant production are generated outside of the solid angle coverage of the GAr, PRISM might provide a further enhancement of the sensitivity.

Second, one can consider DUNE taking data in a beam dump mode, as already done for MiniBooNE. In this case, the neutrino flux, which represents the main source of background to our $Z^\prime$ searches, is significantly reduced. Moreover, the positrons produced as secondary particles in proton interactions would travel through more radiation lengths compared to the target mode due to the significantly deeper depth of the dump, thus increasing the number of $Z^\prime$ produced through resonance.

\section*{Acknowledgements}
We are very grateful to Andrea Celentano and Luc Darme for helpful conversations. The work of FC and IMS is supported by the U.S. Department of
Energy under the award number DE-SC0020250. The work of BD and AT is supported in part by the U.S.~Department of Energy under the award number  DE-SC0010813. 
The work of GG, WJ and JY are supported by the U.S. Department of Energy under the award number DE-SC0011686.

\begin{appendix}

\section{}
Let us assume that $Z^\prime$ decays are isotropic in the center of mass frame, so that we can take the angular distribution of leptons to be $dN_l/d\cos\theta^*=1$, where $\theta^*$ represents the decay angle with respect to the proton beam in this frame. Note that the distribution is normalized to 2 leptons being produced per each decay. The number of leptons per decay being emitted in an angular cone $[\cos\theta_1,\cos\theta_2]$ in the laboratory frame by a $Z^\prime$ with energy $E_{Z^\prime}$ is equal to:
\begin{widetext}
\begin{align}
\begin{split}
n_{l}(E_{Z^\prime},E_l^{\rm th},\theta_{1},\theta_{2})=&\int_{\cos\theta_1}^{\cos\theta_2}d\cos\theta\frac{dN_l}{d\cos\theta^*}\frac{d\cos\theta^*}{d\cos\theta}H[E_l(E_{Z^\prime})-E_l^{\rm th}]
=\\
&\int_{\cos\theta_1}^{\cos\theta_2}d\cos\theta\frac{1 - \frac{p_{Z^\prime}^2}{E_{Z^\prime}^2}}{
		\sqrt{1 - \frac{4 m_l^2}{m_{Z^\prime}^2}} \left(1 - \cos\theta\frac{p_{Z^\prime}}{E_{Z^\prime}}\right)^2}H[E_l(E_{Z^\prime})-E_l^{\rm th}]\,,
\end{split}
\label{number_of_leptons_per_decay}
\end{align}
\end{widetext}
where $E_l^{\rm th}$ is the energy threshold for the detection of lepton $l=e,\mu$, $H$ is the Heaviside function, and $E_l(E_{Z^\prime})$ is the lepton energy in the laboratory frame. The latter is equal to
\begin{equation}
E_l(E_{Z^\prime})=\frac{E_{Z^\prime}}{m_{Z^\prime}}\left (E_l^* + \frac{p_{Z^\prime}}{E_{Z^\prime}} p_{l}^*\cos\theta^*\right)\,,
\label{lepton_energy_lab_frame}
\end{equation}
where $E_l^*=m_{Z^\prime}/2$ and $p_l^*=\sqrt{E_l^{*2}-m_l^2}$ are the energy and the momentum of the lepton in the center of mass frame, respectively. Finally we take 
\begin{eqnarray}
\cos\theta_1&=&\max(\cos\theta_{\rm min}^{\rm exp},\cos\theta_{\rm min}^{\rm kin})\,,\\
\cos\theta_2&=&\min(\cos\theta_{\rm max}^{\rm exp},\cos\theta_{\rm max}^{\rm kin})\,,
\end{eqnarray}
 where $\cos\theta_{\rm min,max}^{\rm exp}$ represent the limits of the angular cone the experiment is sensitive to, whereas $\cos\theta_{\rm min,max}^{\rm kin}$ represent the limits of the angular cone given by the kinematics of the decay.
\end{appendix}

\bibliographystyle{bibi}
\bibliography{Bibliography}

\providecommand{\noopsort}[1]{}\providecommand{\singleletter}[1]{#1}%

\providecommand{\href}[2]{#2}\begingroup\raggedright\begin{thebibliography}{10}

\bibitem{He:1990pn}
X.~G. He, G.~C. Joshi, H.~Lew and R.~R. Volkas, \emph{{NEW Z-prime
  PHENOMENOLOGY}}, \href{https://doi.org/10.1103/PhysRevD.43.R22}{\emph{Phys.
  Rev. D} {\bfseries 43} (1991) 22}.

\bibitem{He:1991qd}
X.-G. He, G.~C. Joshi, H.~Lew and R.~R. Volkas, \emph{{Simplest Z-prime
  model}}, \href{https://doi.org/10.1103/PhysRevD.44.2118}{\emph{Phys. Rev. D}
  {\bfseries 44} (1991) 2118}.

\bibitem{Bauer:2018onh}
M.~Bauer, P.~Foldenauer and J.~Jaeckel, \emph{{Hunting All the Hidden
  Photons}}, \href{https://doi.org/10.1007/JHEP07(2018)094}{\emph{JHEP}
  {\bfseries 07} (2018) 094}
  [\href{https://arxiv.org/abs/1803.05466}{{\ttfamily 1803.05466}}].

\bibitem{Bjorken:1988as}
J.~D. Bjorken, S.~Ecklund, W.~R. Nelson, A.~Abashian, C.~Church, B.~Lu, L.~W.
  Mo, T.~A. Nunamaker and P.~Rassmann, \emph{{Search for Neutral Metastable
  Penetrating Particles Produced in the SLAC Beam Dump}},
  \href{https://doi.org/10.1103/PhysRevD.38.3375}{\emph{Phys. Rev. D}
  {\bfseries 38} (1988) 3375}.

\bibitem{Berryman:2019dme}
J.~M. Berryman, A.~de~Gouvea, P.~J. Fox, B.~J. Kayser, K.~J. Kelly and J.~L.
  Raaf, \emph{{Searches for Decays of New Particles in the DUNE Multi-Purpose
  Near Detector}}, \href{https://doi.org/10.1007/JHEP02(2020)174}{\emph{JHEP}
  {\bfseries 02} (2020) 174}
  [\href{https://arxiv.org/abs/1912.07622}{{\ttfamily 1912.07622}}].

\bibitem{DUNE:2020lwj}
{\scshape DUNE} Collaboration, B.~Abi et~al., \emph{{Deep Underground Neutrino
  Experiment (DUNE), Far Detector Technical Design Report, Volume I
  Introduction to DUNE}},
  \href{https://doi.org/10.1088/1748-0221/15/08/T08008}{\emph{JINST} {\bfseries
  15} (2020) T08008} [\href{https://arxiv.org/abs/2002.02967}{{\ttfamily
  2002.02967}}].

\bibitem{DUNE:2020ypp}
{\scshape DUNE} Collaboration, B.~Abi et~al., \emph{{Deep Underground Neutrino
  Experiment (DUNE), Far Detector Technical Design Report, Volume II: DUNE
  Physics}},  \href{https://arxiv.org/abs/2002.03005}{{\ttfamily 2002.03005}}.

\bibitem{DUNE:2020mra}
{\scshape DUNE} Collaboration, B.~Abi et~al., \emph{{Deep Underground Neutrino
  Experiment (DUNE), Far Detector Technical Design Report, Volume III: DUNE Far
  Detector Technical Coordination}},
  \href{https://doi.org/10.1088/1748-0221/15/08/T08009}{\emph{JINST} {\bfseries
  15} (2020) T08009} [\href{https://arxiv.org/abs/2002.03008}{{\ttfamily
  2002.03008}}].

\bibitem{DUNE:2020txw}
{\scshape DUNE} Collaboration, B.~Abi et~al., \emph{{Deep Underground Neutrino
  Experiment (DUNE), Far Detector Technical Design Report, Volume IV: Far
  Detector Single-phase Technology}},
  \href{https://doi.org/10.1088/1748-0221/15/08/T08010}{\emph{JINST} {\bfseries
  15} (2020) T08010} [\href{https://arxiv.org/abs/2002.03010}{{\ttfamily
  2002.03010}}].

\bibitem{Nardi:2018cxi}
E.~Nardi, C.~D.~R. Carvajal, A.~Ghoshal, D.~Meloni and M.~Raggi,
  \emph{{Resonant production of dark photons in positron beam dump
  experiments}}, \href{https://doi.org/10.1103/PhysRevD.97.095004}{\emph{Phys.
  Rev. D} {\bfseries 97} (2018) 095004}
  [\href{https://arxiv.org/abs/1802.04756}{{\ttfamily 1802.04756}}].

\bibitem{Celentano:2020vtu}
A.~Celentano, L.~Darm\'e, L.~Marsicano and E.~Nardi, \emph{{New production
  channels for light dark matter in hadronic showers}},
  \href{https://doi.org/10.1103/PhysRevD.102.075026}{\emph{Phys. Rev. D}
  {\bfseries 102} (2020) 075026}
  [\href{https://arxiv.org/abs/2006.09419}{{\ttfamily 2006.09419}}].

\bibitem{Aguilar-Arevalo:2018wea}
{\scshape MiniBooNE DM} Collaboration, A.~A. Aguilar-Arevalo et~al.,
  \emph{{Dark Matter Search in Nucleon, Pion, and Electron Channels from a
  Proton Beam Dump with MiniBooNE}},
  \href{https://doi.org/10.1103/PhysRevD.98.112004}{\emph{Phys. Rev. D}
  {\bfseries 98} (2018) 112004}
  [\href{https://arxiv.org/abs/1807.06137}{{\ttfamily 1807.06137}}].

\bibitem{Ballett:2019bgd}
P.~Ballett, T.~Boschi and S.~Pascoli, \emph{{Heavy Neutral Leptons from
  low-scale seesaws at the DUNE Near Detector}},
  \href{https://doi.org/10.1007/JHEP03(2020)111}{\emph{JHEP} {\bfseries 03}
  (2020) 111} [\href{https://arxiv.org/abs/1905.00284}{{\ttfamily
  1905.00284}}].

\bibitem{Coloma:2020lgy}
P.~Coloma, E.~Fern\'andez-Mart\'\i{}nez, M.~Gonz\'alez-L\'opez,
  J.~Hern\'andez-Garc\'\i{}a and Z.~Pavlovic, \emph{{GeV-scale neutrinos:
  interactions with mesons and DUNE sensitivity}},
  \href{https://doi.org/10.1140/epjc/s10052-021-08861-y}{\emph{Eur. Phys. J. C}
  {\bfseries 81} (2021) 78} [\href{https://arxiv.org/abs/2007.03701}{{\ttfamily
  2007.03701}}].

\bibitem{Breitbach:2021gvv}
M.~Breitbach, L.~Buonocore, C.~Frugiuele, J.~Kopp and L.~Mittnacht,
  \emph{{Searching for Physics Beyond the Standard Model in an Off-Axis DUNE
  Near Detector}},  \href{https://arxiv.org/abs/2102.03383}{{\ttfamily
  2102.03383}}.

\bibitem{Atkinson:2021rnp}
M.~Atkinson, P.~Coloma, I.~Martinez-Soler, N.~Rocco and I.~M. Shoemaker,
  \emph{{Heavy Neutrino searches through Double-Bang Events at
  Super-Kamiokande, DUNE, and Hyper-Kamiokande}},
  \href{https://arxiv.org/abs/2105.09357}{{\ttfamily 2105.09357}}.

\bibitem{Schwetz:2021crq}
T.~Schwetz, A.~Zhou and J.-Y. Zhu, \emph{{Constraining active-sterile neutrino
  transition magnetic moments at DUNE near and far detectors}},
  \href{https://arxiv.org/abs/2105.09699}{{\ttfamily 2105.09699}}.

\bibitem{Kelly:2020dda}
K.~J. Kelly, S.~Kumar and Z.~Liu, \emph{{Heavy axion opportunities at the DUNE
  near detector}},
  \href{https://doi.org/10.1103/PhysRevD.103.095002}{\emph{Phys. Rev. D}
  {\bfseries 103} (2021) 095002}
  [\href{https://arxiv.org/abs/2011.05995}{{\ttfamily 2011.05995}}].

\bibitem{Brdar:2020dpr}
V.~Brdar, B.~Dutta, W.~Jang, D.~Kim, I.~M. Shoemaker, Z.~Tabrizi, A.~Thompson
  and J.~Yu, \emph{{Axionlike Particles at Future Neutrino Experiments: Closing
  the Cosmological Triangle}},
  \href{https://doi.org/10.1103/PhysRevLett.126.201801}{\emph{Phys. Rev. Lett.}
  {\bfseries 126} (2021) 201801}
  [\href{https://arxiv.org/abs/2011.07054}{{\ttfamily 2011.07054}}].

\bibitem{Batell:2009di}
B.~Batell, M.~Pospelov and A.~Ritz, \emph{{Exploring Portals to a Hidden Sector
  Through Fixed Targets}},
  \href{https://doi.org/10.1103/PhysRevD.80.095024}{\emph{Phys. Rev. D}
  {\bfseries 80} (2009) 095024}
  [\href{https://arxiv.org/abs/0906.5614}{{\ttfamily 0906.5614}}].

\bibitem{deNiverville:2012ij}
P.~deNiverville, D.~McKeen and A.~Ritz, \emph{{Signatures of sub-GeV dark
  matter beams at neutrino experiments}},
  \href{https://doi.org/10.1103/PhysRevD.86.035022}{\emph{Phys. Rev. D}
  {\bfseries 86} (2012) 035022}
  [\href{https://arxiv.org/abs/1205.3499}{{\ttfamily 1205.3499}}].

\bibitem{Coloma:2015pih}
P.~Coloma, B.~A. Dobrescu, C.~Frugiuele and R.~Harnik, \emph{{Dark matter beams
  at LBNF}}, \href{https://doi.org/10.1007/JHEP04(2016)047}{\emph{JHEP}
  {\bfseries 04} (2016) 047}
  [\href{https://arxiv.org/abs/1512.03852}{{\ttfamily 1512.03852}}].

\bibitem{deNiverville:2018dbu}
P.~deNiverville and C.~Frugiuele, \emph{{Hunting sub-GeV dark matter with the
  NO$\nu$A near detector}},
  \href{https://doi.org/10.1103/PhysRevD.99.051701}{\emph{Phys. Rev. D}
  {\bfseries 99} (2019) 051701}
  [\href{https://arxiv.org/abs/1807.06501}{{\ttfamily 1807.06501}}].

\bibitem{Jordan:2018qiy}
J.~R. Jordan, Y.~Kahn, G.~Krnjaic, M.~Moschella and J.~Spitz, \emph{{Severe
  Constraints on New Physics Explanations of the MiniBooNE Excess}},
  \href{https://doi.org/10.1103/PhysRevLett.122.081801}{\emph{Phys. Rev. Lett.}
  {\bfseries 122} (2019) 081801}
  [\href{https://arxiv.org/abs/1810.07185}{{\ttfamily 1810.07185}}].

\bibitem{deGouvea:2018cfv}
A.~de~Gouv\^ea, P.~J. Fox, R.~Harnik, K.~J. Kelly and Y.~Zhang, \emph{{Dark
  Tridents at Off-Axis Liquid Argon Neutrino Detectors}},
  \href{https://doi.org/10.1007/JHEP01(2019)001}{\emph{JHEP} {\bfseries 01}
  (2019) 001} [\href{https://arxiv.org/abs/1809.06388}{{\ttfamily
  1809.06388}}].

\bibitem{DeRomeri:2019kic}
V.~De~Romeri, K.~J. Kelly and P.~A.~N. Machado, \emph{{DUNE-PRISM Sensitivity
  to Light Dark Matter}},
  \href{https://doi.org/10.1103/PhysRevD.100.095010}{\emph{Phys. Rev. D}
  {\bfseries 100} (2019) 095010}
  [\href{https://arxiv.org/abs/1903.10505}{{\ttfamily 1903.10505}}].

\bibitem{Magill:2018tbb}
G.~Magill, R.~Plestid, M.~Pospelov and Y.-D. Tsai, \emph{{Millicharged
  particles in neutrino experiments}},
  \href{https://doi.org/10.1103/PhysRevLett.122.071801}{\emph{Phys. Rev. Lett.}
  {\bfseries 122} (2019) 071801}
  [\href{https://arxiv.org/abs/1806.03310}{{\ttfamily 1806.03310}}].

\bibitem{Ballett:2019xoj}
P.~Ballett, M.~Hostert, S.~Pascoli, Y.~F. Perez-Gonzalez, Z.~Tabrizi and
  R.~Zukanovich~Funchal, \emph{{$Z^\prime$s in neutrino scattering at DUNE}},
  \href{https://doi.org/10.1103/PhysRevD.100.055012}{\emph{Phys. Rev. D}
  {\bfseries 100} (2019) 055012}
  [\href{https://arxiv.org/abs/1902.08579}{{\ttfamily 1902.08579}}].

\bibitem{Dev:2021xzd}
P.~S.~B. Dev, D.~Kim, K.~Sinha and Y.~Zhang, \emph{{New Interference Effects
  from Light Gauge Bosons in Neutrino-Electron Scattering}},
  \href{https://arxiv.org/abs/2105.09309}{{\ttfamily 2105.09309}}.

\bibitem{GEANT4:2002zbu}
{\scshape GEANT4} Collaboration, S.~Agostinelli et~al., \emph{{GEANT4--a
  simulation toolkit}},
  \href{https://doi.org/10.1016/S0168-9002(03)01368-8}{\emph{Nucl. Instrum.
  Meth. A} {\bfseries 506} (2003) 250}.

\bibitem{Allison:2006ve}
J.~Allison et~al., \emph{{Geant4 developments and applications}},
  \href{https://doi.org/10.1109/TNS.2006.869826}{\emph{IEEE Trans. Nucl. Sci.}
  {\bfseries 53} (2006) 270}.

\bibitem{Allison:2016lfl}
J.~Allison et~al., \emph{{Recent developments in Geant4}},
  \href{https://doi.org/10.1016/j.nima.2016.06.125}{\emph{Nucl. Instrum. Meth.
  A} {\bfseries 835} (2016) 186}.

\bibitem{PhysRevD.98.112004}
{\scshape The MiniBooNE-DM Collaboration} Collaboration, A.~A. Aguilar-Arevalo,
  M.~Backfish, A.~Bashyal, B.~Batell, B.~C. Brown, R.~Carr, A.~Chatterjee,
  R.~L. Cooper, P.~deNiverville, R.~Dharmapalan, Z.~Djurcic, R.~Ford, F.~G.
  Garcia, G.~T. Garvey, J.~Grange, J.~A. Green, E.-C. Huang, W.~Huelsnitz,
  I.~L. de~Icaza~Astiz, G.~Karagiorgi, T.~Katori, W.~Ketchum, T.~Kobilarcik,
  Q.~Liu, W.~C. Louis, W.~Marsh, C.~D. Moore, G.~B. Mills, J.~Mirabal,
  P.~Nienaber, Z.~Pavlovic, D.~Perevalov, H.~Ray, B.~P. Roe, M.~H. Shaevitz,
  S.~Shahsavarani, I.~Stancu, R.~Tayloe, C.~Taylor, R.~T. Thornton, R.~G.
  Van~de Water, W.~Wester, D.~H. White and J.~Yu, \emph{Dark matter search in
  nucleon, pion, and electron channels from a proton beam dump with miniboone},
  \href{https://doi.org/10.1103/PhysRevD.98.112004}{\emph{Phys. Rev. D}
  {\bfseries 98} (2018) 112004}.

\bibitem{Papadimitriou:2017zai}
{\scshape DUNE} Collaboration, V.~Papadimitriou, \emph{{Design of the LBNF
  Beamline}},  in \emph{{38th International Conference on High Energy
  Physics}}, 2017.

\bibitem{Zyla:2020zbs}
{\scshape Particle Data Group} Collaboration, P.~Zyla et~al., \emph{{Review of
  Particle Physics}}, \href{https://doi.org/10.1093/ptep/ptaa104}{\emph{PTEP}
  {\bfseries 2020} (2020) 083C01}.

\bibitem{Dutta:2020vop}
B.~Dutta, D.~Kim, S.~Liao, J.-C. Park, S.~Shin, L.~E. Strigari and A.~Thompson,
  \emph{{Searching for Dark Matter Signals in Timing Spectra at Neutrino
  Experiments}},  \href{https://arxiv.org/abs/2006.09386}{{\ttfamily
  2006.09386}}.

\bibitem{Croon:2020lrf}
D.~Croon, G.~Elor, R.~K. Leane and S.~D. McDermott, \emph{{Supernova Muons: New
  Constraints on $Z$' Bosons, Axions and ALPs}},
  \href{https://doi.org/10.1007/JHEP01(2021)107}{\emph{JHEP} {\bfseries 01}
  (2021) 107} [\href{https://arxiv.org/abs/2006.13942}{{\ttfamily
  2006.13942}}].

\bibitem{Knapen:2017xzo}
S.~Knapen, T.~Lin and K.~M. Zurek, \emph{{Light Dark Matter: Models and
  Constraints}}, \href{https://doi.org/10.1103/PhysRevD.96.115021}{\emph{Phys.
  Rev. D} {\bfseries 96} (2017) 115021}
  [\href{https://arxiv.org/abs/1709.07882}{{\ttfamily 1709.07882}}].

\bibitem{Escudero:2019gzq}
M.~Escudero, D.~Hooper, G.~Krnjaic and M.~Pierre, \emph{{Cosmology with A Very
  Light L$_{\mu}$ \ensuremath{-} L$_{\tau}$ Gauge Boson}},
  \href{https://doi.org/10.1007/JHEP03(2019)071}{\emph{JHEP} {\bfseries 03}
  (2019) 071} [\href{https://arxiv.org/abs/1901.02010}{{\ttfamily
  1901.02010}}].

\bibitem{MicroBooNE:2019rgx}
{\scshape MicroBooNE} Collaboration, C.~Adams et~al., \emph{{Reconstruction and
  Measurement of $\mathcal{O}$(100) MeV Energy Electromagnetic Activity from
  $\pi^0 \rightarrow \gamma\gamma$ Decays in the MicroBooNE LArTPC}},
  \href{https://doi.org/10.1088/1748-0221/15/02/P02007}{\emph{JINST} {\bfseries
  15} (2020) P02007} [\href{https://arxiv.org/abs/1910.02166}{{\ttfamily
  1910.02166}}].

\bibitem{Davier:1989wz}
M.~Davier and H.~Nguyen~Ngoc, \emph{{An Unambiguous Search for a Light Higgs
  Boson}}, \href{https://doi.org/10.1016/0370-2693(89)90174-3}{\emph{Phys.
  Lett. B} {\bfseries 229} (1989) 150}.

\bibitem{Bjorken:2009mm}
J.~D. Bjorken, R.~Essig, P.~Schuster and N.~Toro, \emph{{New Fixed-Target
  Experiments to Search for Dark Gauge Forces}},
  \href{https://doi.org/10.1103/PhysRevD.80.075018}{\emph{Phys. Rev. D}
  {\bfseries 80} (2009) 075018}
  [\href{https://arxiv.org/abs/0906.0580}{{\ttfamily 0906.0580}}].

\bibitem{Andreas:2012mt}
S.~Andreas, C.~Niebuhr and A.~Ringwald, \emph{{New Limits on Hidden Photons
  from Past Electron Beam Dumps}},
  \href{https://doi.org/10.1103/PhysRevD.86.095019}{\emph{Phys. Rev. D}
  {\bfseries 86} (2012) 095019}
  [\href{https://arxiv.org/abs/1209.6083}{{\ttfamily 1209.6083}}].

\bibitem{Ilten:2018crw}
P.~Ilten, Y.~Soreq, M.~Williams and W.~Xue, \emph{{Serendipity in dark photon
  searches}}, \href{https://doi.org/10.1007/JHEP06(2018)004}{\emph{JHEP}
  {\bfseries 06} (2018) 004}
  [\href{https://arxiv.org/abs/1801.04847}{{\ttfamily 1801.04847}}].

\bibitem{Planck:2018vyg}
{\scshape Planck} Collaboration, N.~Aghanim et~al., \emph{{Planck 2018 results.
  VI. Cosmological parameters}},
  \href{https://doi.org/10.1051/0004-6361/201833910}{\emph{Astron. Astrophys.}
  {\bfseries 641} (2020) A6}
  [\href{https://arxiv.org/abs/1807.06209}{{\ttfamily 1807.06209}}].

\bibitem{Nelson:2008tn}
A.~E. Nelson and J.~Walsh, \emph{{Chameleon vector bosons}},
  \href{https://doi.org/10.1103/PhysRevD.77.095006}{\emph{Phys. Rev. D}
  {\bfseries 77} (2008) 095006}
  [\href{https://arxiv.org/abs/0802.0762}{{\ttfamily 0802.0762}}].

\bibitem{Depta:2020wmr}
P.~F. Depta, M.~Hufnagel and K.~Schmidt-Hoberg, \emph{{Robust cosmological
  constraints on axion-like particles}},
  \href{https://doi.org/10.1088/1475-7516/2020/05/009}{\emph{JCAP} {\bfseries
  05} (2020) 009} [\href{https://arxiv.org/abs/2002.08370}{{\ttfamily
  2002.08370}}].

\bibitem{DUNE:2021tad}
{\scshape DUNE} Collaboration, A.~Abed~Abud et~al., \emph{{Deep Underground
  Neutrino Experiment (DUNE) Near Detector Conceptual Design Report}},
  \href{https://arxiv.org/abs/2103.13910}{{\ttfamily 2103.13910}}.

\end{thebibliography}\endgroup

\end{document}